# Strengthening Infrastructure Resilience to Hurricanes by Modeling Transportation and Electric Power Network Interdependencies

**Tasnuba Binte Jamal**
Department of Civil, Environmental and Construction Engineering
University of Central Florida
Orlando, FL 32816
Email: tasnubabinte.jamal@ucf.edu

**Samiul Hasan, Ph.D. (Corresponding Author)**
Department of Civil, Environmental and Construction Engineering
University of Central Florida
Orlando, FL 32816
Email: samiul.hasan@ucf.edu

**Omar I. Abdul-Aziz, Ph.D.**
Department of Civil and Environmental Engineering
West Virginia University
Morgantown, West Virginia
Email: oiabdulaziz@mail.wvu.edu

**Pallab Mozumder, Ph.D.**
Department of Earth & Environment and Department of Economics
Florida International University
Miami, Florida.
Email: mozumder@fiu.edu

**Rounak Meyur, Ph.D.**
Data Scientist
Data Science and Machine Intelligence Group
Pacific Northwest National Laboratory
Email: rounak.meyur@pnnl.gov

## ABSTRACT

Community resilience is significantly affected by infrastructure disruptions during hurricanes. Resilience is generally defined as the ability of a system to manage shocks and return to a normal state in response to an extreme event. Due to the interconnected and interdependent relationships among infrastructure systems, the restoration process of a system is further delayed when other systems are disrupted. This study presents an agent-based model (ABM) developed to simulate the resilience of infrastructures to hurricanes, focusing on the interdependencies between electric power and transportation networks. To study infrastructure resilience to a hurricane, a library of agents has been created including electric power network, transportation network, wind/flooding hazards, and household agents. The ABM is applied over the households in ZIP Code 33147 of Miami-Dade County, Florida and infrastructures supporting these households. Interdependencies between the two networks are modeled in two ways, (i) representing the role of transportation in fuel delivery to power plants and restoration teams' access to failed power system components, (ii) impact of power outage on transportation network components. We simulate three restoration strategies: component based, distance based, and traffic-light based restoration. The model is validated against Hurricane Irma data, showing consistent behavior with varying hazard intensities. Scenario analyses reveal the impact of restoration strategies, road accessibility, and wind speed on power service restoration. Results demonstrate that a traffic-light based restoration strategy efficiently prioritizes signal recovery without delaying household power restoration time. Restoration of power services will be faster if fuel transportation to power plants and restoration efforts are not delayed by inaccessible roads due to flooding. The developed agent-based model can be used as a decision-support tool by policy makers and utility/emergency managers in evaluating power outage restoration strategies using available resources.

## INTRODUCTION

The frequency and intensity of hurricanes have increased recently due to climate change and global warming. During hurricanes, sustained winds and severe flooding cause significant infrastructure disruptions including power outages, disturbances in water supply and wastewater systems, telecommunication failures, and transportation system disruptions. Resilience against such infrastructure disruptions should be an important aspect of community resilience where a resilient infrastructure system would be able to quickly return to its typical condition following a disruption (Hosseini et al. 2016; Roy et al. 2019). Resilience is characterized as the ability of a disaster-affected community to return to normal life. At the outset of a hurricane, people begin to experience difficulty due to infrastructure disruptions; after a certain period, the maximum population experiencing hardship is reached, and then, the population begins to recover. A resilient community is one that recovers to having a similar or higher level of service in a reasonable time following a disaster, with no significant degradation of key services. One way to enhance community resilience is to strengthen its critical lifelines (power, water, transportation, internet) services (Bruneau et al. 2003) and capital—to reduce the likelihood of damage or loss through mitigation. This paper aims to develop a resilience modeling framework of hurricane-affected communities considering multiple critical lifelines.

A community is composed of different sectors, including residential, educational, business, and healthcare that are supported by a set of infrastructures such as power, water, transportation, and telecommunication systems. The residential sector is a crucial component of every community. It has a direct impact on individuals, shapes the wellbeing of a community, and represents a significant portion of the building market. As a result, restoration of infrastructure services of the affected households is critical to the rebuilding of a community.

To have a holistic understanding of community resilience, it is important to study different community sectors under infrastructure disruptions. Prior studies (Esmalian et al. 2022; Ouyang and Dueñas-Osorio 2014) proposed frameworks to study household resilience after hurricanes by quantifying experienced hardship due to disruptions in power networks. Prior studies also investigated disruptions to transportation networks and their impacts to healthcare sector (Dong et al. 2020) and disruptions in water distribution networks and their impacts to individuals (Cimellaro et al. 2016). These studies investigated disruptions to a single critical component of the corresponding infrastructure system. However, infrastructure systems are highly interconnected and interdependent (Grafius et al. 2020; Hasan and Foliente 2015; Rinaldi et al. 2001). Two infrastructure systems are interdependent when each depends on the other (Rinaldi et al. 2001), or when the functioning of one infrastructure system is required for the functionality of the other (Zimmerman 2001). For instance, damages in energy stations caused by Hurricane Sandy had a substantial impact on the operations of transportation facilities (Haraguchi and Kim 2016). Due to the complex and interconnected relationships among infrastructure systems, the restoration process of a system is further slowed when other systems are affected (Guidotti et al. 2016). For instance, power outages fail traffic signals and water supply pumps. Debris-covered roadways are inaccessible to emergency response and recovery teams (Bigger E. John et al. 2009; Rinaldi et al. 2001). Consequently, infrastructure and emergency services become unavailable, and the quality of life for the population degrades. As such, focus should be given to modeling the interdependency among infrastructure systems.

Several studies investigated interdependencies among multiple infrastructure systems (e.g., water supply, electric power) and their impacts on community sectors (e.g., households, schools, businesses). For example, previous studies investigated the effect of disruptions in water supply network along with electric power network on residential, commercial, industrial, healthcare sectors due to an earthquake (Guidotti et al. 2016; Hassan and Mahmoud 2021; Nasrazadani and Mahsuli 2020); on educational sector (Aghababaei and Koliou 2022b) and education, business, and healthcare sectors due to a tornado (Aghababaei and Koliou 2022a) and a hurricane (Mühlhofer et al. 2023). Hassan and Mahmoud (2021) studied the impacts of building damages along with supporting infrastructure on healthcare and education sectors due to an earthquake considering that water supply depends on power networks. Yu and Baroud (2020) also studied interdependent water and power networks after an earthquake. Nofal et al. (2024) introduced a building-level post-hazard functionality model considering interdependencies between population, buildings, and multiple infrastructures for flood-exposed communities. Although many infrastructure systems (e.g., electric power, water supply) rely on a transportation network, the role of its functionality to the recovery of other infrastructures has been understudied in literature. By simulating household impacts of hurricanes in conjunction with interdependent components of transportation and other infrastructure systems within a comprehensive framework, it is possible to evaluate infrastructure resilience more accurately.

Previous studies proposed strategies towards offering equitable power service across different group of people (race, age, mobility) after hurricanes (Esmalian et al. 2022) and how resilience of power service decreases because of its dependence on water infrastructures after tornados (Aghababaei and Koliou 2022a; b). On the other hand, we add a new dimension by prioritizing the restoration of both power and transportation infrastructures as equally important after considering interdependency issues. We evaluate different restoration strategies so that traffic signals receive power without significantly delaying the restoration time of household power services after hurricanes, considering the interdependent relationships between power and transportation infrastructure systems.

Resilience usually indicates the ability of a system to return to its normal state or situation after a disruption due to an extreme event (Hosseini et al. 2016; Roy et al. 2019). At the onset of a hurricane, system performance starts to decrease in the affected area, reaches maximum drop after a certain time, and then starts to recover. Since system performance is not at usual level, the amount of decrease in system performance compared to regular periods can be termed as the loss of resilience in a disaster affected region (Roy et al. 2019). Bruneau et al. (2003) developed a framework for measuring loss of resilience considering four dimensions: robustness, rapidity, resourcefulness, and redundancy. They also proposed an equation to measure resilience loss due to an earthquake. Later, this equation was applied to many other contexts such as for hurricanes (Roy et al. 2019) and tornados (Aghababaei and Koliou 2022a; b). Cimellaro et al. (2010) also provided another framework to quantify resilience using an analytical function that may fit both technical and organizational issues. Resilience has been studied on a wide range of sectors for different types of natural hazards such as resilience of power and water infrastructures due to tornados (Aghababaei and Koliou 2022a; b), resilience of bridges and embankments due to earthquake (Akiyama et al. 2020; Argyroudis and Mitoulis 2021; Forcellini 2023), flood (Argyroudis and Mitoulis 2021), tsunami and corrosion (Akiyama et al. 2020). For hurricanes, resilience has been studied for human mobility (Roy et al. 2019), population activity (Jamal and Hasan 2023a), structural buildings (Abdelhady et al. 2020). Infrastructure resilience under hurricanes has also been investigated previously. For example, previous studies investigated the resilience of power systems after Hurricane Sandy (Henry and Ramirez-Marquez 2016) and Hurricane Ike (Ouyang and Dueñas-Osorio 2014), and the importance of microgrids to ensure the functionality of power systems during hurricanes (Aros-Vera et al. 2021). Our study focuses on improving resilience of multiple infrastructure systems due to a hurricane considering the interdependent relationships between them.

This research develops an Agent Based Model (ABM) to advance our understanding of the impacts of disruptions on infrastructure resilience during and after hurricanes. ABM is suitable for simulating the complexity of interconnected infrastructures (Ouyang 2014), since it aggregates simple micro-level interactions among agents into complex macro-level reactions of a community. Previous studies utilized ABM to study disaster resilience and recovery of communities. Using ABM, Eid and El-adaway (2017) developed a comprehensive decision-making framework to simulate the recovery progress of hurricane affected communities. They considered the recovery of affected residents' wealth, social vulnerabilities, government aid, and accessible insurance plans. Nejat and Damnjanovic (2012) employed ABM to model the dynamic interactions between homeowners and their neighbors and their effects on a homeowner's reconstruction decision in the recovery process. Moradi and Nejat (2020) proposed a spatial ABM to investigate the recovery from housing damages and found that combination of internal (income, education, race, physical damage), interactive (neighbors' recovery), and external (financial assistance, infrastructure recovery, and community asset recovery) drivers influence a household's recovery decisions. Prior studies also showed that ABM can effectively model the behavior of one system like electric power network (Esmalian et al. 2022) and multiple complex systems (electric power network and water supply network) including their dependent nature (Aghababaei and Koliou 2022a; b). Prior ABM studies (Aghababaei and Koliou 2022a; b; Nasrazadani and Mahsuli 2020) considered that the water supply network depends on having power access through the electric power network, and households, schools, business, and healthcare facilities depend on both the electric power and water supply network to operate. As such, ABM is a reliable method to simulate how infrastructure systems are disrupted due to a hurricane and to evaluate the effect of possible restoration strategies.

Using an ABM framework, this study quantifies the loss in infrastructure resilience to hurricanes considering interdependent relationships between infrastructure systems. It also evaluates the effect of different restoration strategies as post-disaster actions to enhance their resilience. This study contributes to the literature by answering the following research questions:

1. *How do the disruptions in a transportation network delay the restoration process in an electric power network and increase the overall "loss of infrastructure resilience" due to a hurricane?*

   If power plants heavily rely on fuel transportation and flooded roads cannot be used by restoration teams, restoration process for different components of the power network is delayed and customers face longer power outage. We measure the increase in restoration time and loss of resilience of the households to power service because of the interdependent relationship between electric power network and transportation network.

2. *What are the impacts of different restoration strategies to strengthen resilience of power and transportation systems to a hurricane?*

   Power outage duration and number of households without power can vary depending on the adopted restoration strategies. We implemented one restoration strategy from existing literature and propose two alternative restoration strategies and evaluate their performances.

## STUDY AREA

For this paper, we select a ZIP Code in Miami-Dade County of Florida, USA as our study area. We simulate the scenario for Hurricane Irma and a higher intensity hurricane in the same area. Although Hurricane Irma hit Monroe County of Florida, we did not focus on Monroe County because the population density of this county is very small, and required data to generate synthetic power network was not available for this county. The variation of wind speed caused by a hurricane is small in a county depending on the area of the county (e.g., in Hurricane Irma, wind speed variation in Miami-Dade County was 5 mph). We focused on ZIP Code 33147 of Miami-Dade County because of two reasons: (i) this is a highly residential area with 80% residents, and (ii) the predicted runoff depth varies over the block-groups present in it. **Figure 1(a)** shows the runoff depth over Miami-Dade County – the runoff depth of ZIP Code 33147 ranges from 13 to 26 inches. **Figure 1(b)** shows the spatial distribution of the building/property inventory of ZIP Code 33147. We also developed an ABM to understand evacuation behavior for the same ZIP Code in a prior study (Roy et al. 2022).

## METHODOLOGY

### Overview of the Agent Based Model (ABM)

**Figure 2** presents the proposed ABM framework which captures the dynamic processes that cause households' power interruptions. The hazard component, represented by its severity, affects the infrastructure systems. The electrical and transportation infrastructure systems include component fragility and network topology. When the components of an infrastructure system are more fragile, the likelihood of severe failure increases. In addition, the network topology influences a power system through connectivity to power generator. Geographical location, elevation, and drainage system influences accessibility of the transportation network. Repairing resource availability also significantly influences the outage duration of the affected community. The details of the agent and the interactions among them are described in the following sections.

## Model Implementation

**Figure 3** presents the unified modeling language class diagram of the agent-based model (ABM). It summarizes each agent's attributes and their operating methods. In this figure, the letter (*a*) stands for attributes and the letter (*m*) stands for methods. The functionality of an agent or a component depends on the functionality of other agents/components in the system. The interdependent relationships among different agents are demonstrated by arrows and colors. The color of an arrow indicates from which agents they are receiving inputs. Besides the relationship of an agent with other agents, various components of agents (e.g., transmission and distribution components of power network) interact among themselves. This intra-relationship is described in detail in the power network agent section. The integration of the mentioned attributes and methods of the agents enables simulating the extent of infrastructure failures by wind and flooding, and their restoration processes.

The failure and restoration of the agents were implemented in Repast Simphony 2.9 (North et al. 2013) which is an agent-based modeling framework. The simulation followed discrete time-steps; at each step, time was increased by an hour. The simulations were run until all failed power components were restored. To capture stochasticity, Monte Carlo experimentation was used. Since one of the objectives of the model was to determine the percentage of the households with power service, simulation experiments were replicated as many times as the mean value of the proportion of households with power service achieved a 90% confidence interval with 10% error (Hanh 1972). All figures and tables in the result section are the average of these simulation runs. Resilience of households' power services is estimated for different hazard severity and restoration strategies. Details of different methods and agent attributes are described below.

## Hazard Agent

This study considered wind and flood hazards. The ABM wind agent causes the failure of the power network agent, and the flood agent disrupts the links and intersections of the transportation network. Besides wind and flood hazards, the power network and transportation network can also be damaged by treefall-induced debris. Here, we are considering only wind-induced power component failure and flood-induced transportation network disruptions due to available literature on wind and flood-induced failures. Prior studies also reported that most hurricane damages are caused by wind (Panteli et al. 2017).

The HAZUS-MH wind model was used to estimate the wind speed for Hurricane Irma (Vickery et al. 2000, 2006). This model generates the wind speed profile due to a hurricane in a probabilistic manner. Using this model, we determined the maximum sustained wind speed and 5-s gust speed at the census tract level in relation to their distance from the hurricane's center. To predict the rainfall scenarios of flood/runoff depths throughout the basin, we ran the model developed by Huq and Abdul-Aziz (2021), at the hourly time-step for 10, 20, 30, and 40 inch uniform (in time and space) rainfall scenarios over 24 hours. The runoff depths were then extracted at the resolution of 1 km x 1 km blocks for Miami-Dade County (**Figure 1(a)**).

This study assumed that restoration personnel cannot use flooded road links. When flood/runoff water drains into canals or lakes, road linkages resume functioning. Based on our understanding and experience, a 12-inch runoff in these highly paved and channelized areas (essentially Florida International University campus neighborhoods) would take 16 to 24 hours to completely drain into the neighboring canals (urban streams) and lakes, which eventually carry stormwater into the sea. This range of drainage time yields a drainage rate of 0.50 to 0.75 inches per hour. Note that the drainage rate can vary from the beginning to the end of the storm. Here, we used an average (0.65 inch/hour) drainage rate. It is important to mention that we used surface water runoff drainage rate (0.50 to 0.75 inch/hour) into the canals and lakes – and not the soil infiltration rate, which is a completely different parameter and is usually much (an order of magnitude) slower.

**Electric Power Network Agent**

The electric power network is considered as a connected grid system. It consists of power plants, transmission towers, transmission lines, substations, distribution poles, and conductors. Electricity passes from a power plant through high voltage transmission lines to a substation; then electricity flows from a substation to households through medium and low voltage distribution lines (**Figure 4**).

We collected the actual data for power plants, substations, transmission towers and lines from U.S. Energy Information Administration (EIA) (atlas.eia.gov) and Homeland Infrastructure Foundation-Level Data (HIFLD) (hifld-geoplatform.opendata.arcgis.com/). However, the distribution lines from substation to households are unavailable for proprietary and security reasons. A synthetic power distribution network (**Figure 5**) was created for the ZIP Code 33147 of Miami-Dade County using the methods developed by Meyur et al. (2020). The synthetic power network is generated to represent the topology of the actual network by solving an optimization algorithm that minimizes the network's overall length, given structural and power flow constraints. The algorithm uses coordinates of households and substations, average hourly electricity demand of households, and road networks and it determines the coordinates of the synthetized distribution systems (e.g., poles, conductors) of the power network. According to the EIA data, the study area's power plant burns biomass. Since urban woods are a widely used source of biomass, we assumed the nearest source of urban wood as the fuel source, located 17.5 km from the power plant. The entry point, connecting the fuel source to the nearest road network, is located at latitude = 25.76105 and longitude = -80.43431.

*Power Component Failure*

Fragility curves are usually utilized to estimate infrastructure damage from natural disasters (Esmalian et al. 2022; Ouyang and Dueñas-Osorio 2014; Winkler et al. 2010). Fragility curves calculate the probability of the failure of a structural component (i.e. P[wind]) based on wind speed. In each iteration, the failure probability would be compared against a random variable drawn from a uniform distribution ($r \in [0,1]$). If the failure probability exceeds the generated random number (r), a power system component will fail (**Figure 6**). This model considers failure in power network for the following components: substations, transmission support structures, transmission lines, distribution poles, and conductors (Esmalian et al. 2022; Mensah and Dueñas-Osorio 2016; Ouyang and Dueñas-Osorio 2014). As hurricanes seldom damage power plants, failures in power plants were not considered in this study (Esmalian et al. 2022; Ouyang and Dueñas-Osorio 2014).

*Substation Agent*

Fragility functions provide failure probability depending on exposed wind speed, terrain, and structural characteristics. The failure probability of a substation ($P_{f,sub,j}$) can be defined by the following lognormal fragility function (Ouyang and Dueñas-Osorio 2014):

$$P_{f,sub,j}\left(D \geq d_j \middle| w = x\right) = \int_0^x \frac{1}{w\sigma_j\sqrt{2\pi}} exp\left(-\frac{\left(ln\,(w) - \mu_j\right)^2}{2\sigma_j^2}\right) dw \tag{1}$$

For a given substation, these curves generate four probabilities ($P_{f,sub,low} > P_{f,sub,moderate} > P_{f,sub,severe} > P_{f,sub,complete}$) for the substation damage D exceeding a certain level $d_{i,j}$ where $j$ represents the damage state of the substation = {low, moderate, severe, complete}. These levels correspond to the extent of damage a substation might experience under certain wind speeds. The variable w represents wind speed and x is a specific wind speed value, and $\mu_j$ and $\sigma_j$ are used to define the lognormal fragility curve. These values of the fragility curves for each damage level, and each type of modeled terrain and building type are taken from HAZUS-MH4 (FEMA 2008).

For each substation and each possible damage state (low, moderate, severe, complete), the probabilities of exceeding these damage states are calculated using the lognormal fragility function. This results in a set of cumulative probabilities for each substation. To determine the actual damage state of a substation, a uniformly distributed random variable $U$ within the range [0, 1] is generated and compared with the cumulative probabilities of the damage states. For example, let the cumulative probabilities of exceeding low, moderate, severe, and complete damage states for a specific substation be $P_{sub,low}$, $P_{sub,moderate}$, $P_{sub,severe}$, and $P_{sub,complete}$. If $U$ is less than $P_{sub,low}$, then the substation is assigned a damage state of low. If $P_{f,sub,low} < U < P_{f,sub,moderate}$, then the substation is assigned a damage state of moderate. Similarly, severe and complete damage states were assigned. Following a previous study (Esmalian et al. 2022), we assumed that except for low damage state, other damage states yield a functionality failure and loss of connectivity for the cascading failure analysis.

To repair a substation agent, restoration time in hours (h) is assigned following normal distributions, *N (mean, standard deviation)* such as for moderate: $(72\ h, 36\ h)$, severe: $(168\ h, 84\ h)$, and complete: $(720\ h,\ 360\ h)$ damage levels. Required numbers of restoration team are considered as 6, 14, and 60 for moderate, severe, and complete damage levels, respectively following previous studies (Esmalian et al. 2022; Ouyang and Dueñas-Osorio 2014).

*Transmission Component Agents*

We obtained failure probability of transmission tower agent from an equation (**Equation 2**) developed by Quanta Technology (2009); this equation was used before for hurricanes by Mensah and Dueñas-Osorio (2016); Ouyang and Dueñas-Osorio (2014).

$$P_{f,TT} = 2\ \times 10^{-7} e^{0.0834x} \tag{1}$$

To determine the probability of the failure of a transmission line agent we used **Equation 3**, adopted by Panteli et al. (2017) and Esmalian et al. (2022).

$$P_{f,TL} = \begin{cases} 0.01, & if\ x < w_{critical} \\ TL, & if\ w_{critical} < x \leq w_{collapse} \\ 1, & x > w_{collapse} \end{cases} \tag{2}$$

In **Equation 3**, when the wind speed $(x)$ falls below a certain threshold, the likelihood for a transmission line failure is low (0.01). Here, $w_{collapse}$ denotes a scenario where the survival probability of the transmission line is extremely low. The probability of failure of the component (T$L$) is determined by establishing a linear relationship between $w_{critical}$ and $w_{collapse}$. These thresholds are assumed to range from 30 to 60 m/s based on empirical research (Esmalian et al. 2022; Panteli et al. 2017).

To repair a transmission agent, following previous studies (Esmalian et al. 2022; Mensah and Dueñas-Osorio 2016; Ouyang and Dueñas-Osorio 2014), restoration time in hours (h) is assigned as normal distributions for towers: $(72\ h, 36\ h)$ and transmission lines: $(48\ h, 24\ h)$.

*Distribution Components*

Distribution components of the synthetic electric power network consist of distribution poles and conductors (Esmalian et al. 2022; Ouyang and Dueñas-Osorio 2014). We used failure probability functions (**Equations 4 and 5)** developed by Quanata Technology (2008); Quanta Technology (2009) and adopted in previous studies (Esmalian et al. 2022; Mensah and Dueñas-Osorio 2016; Ouyang and Dueñas-Osorio 2014).

$$P_{f,pole} = 0.0001 \times e^{0.0421x} \quad (3)$$

$$P_{f,conductors} = 8 \times 10^{-12} \times x^{5.1731} \quad (4)$$

To repair a distribution agent, restoration time in hours (h) is assigned as normal distributions for distribution poles: $(5\ h,\ 2.5\ h)$ and conductors: $(4\ h, 2\ h)$. For restoration time of all types of power system components, we replaced the negative value (if generated) by a small value 0.0833 (i.e. 5 minutes).

The required numbers of restoration team are assumed as 6, 4, and 1 for transmission tower, line, and distribution agents, respectively (Esmalian et al. 2022; Ouyang and Dueñas-Osorio 2014; Quanata Technology 2008). The total number of agents of the electric power network is 6451. In **Equations 2 - 5**, the variable $x$ represents the exposed wind speed in miles per hour.

*Modeling Cascading Failures*

When a component of an electric power network fails, it propagates through the network to customer ends, resulting in connectivity loss and known as cascading failures (Aghababaei and Koliou 2022a; Winkler et al. 2010). The proposed model considered cascading failures due to the intra-dependencies among electric power network agents. If there is any failed component from the power generator along the path towards a household, it would be out of electricity. Each iteration of the ABM assessed the connectivity of the households to a power generator to capture cascading failure.

**Transportation Network Agent**

Transportation network agents consist of road links, intersections, and traffic lights. Since transportation infrastructure components are less likely to be damaged by hurricane-induced winds, such damages were not considered in this study. We considered that power restoration teams cannot use roadways inundated by hurricane-induced runoff (Bigger E. John et al. 2009; Rinaldi et al. 2001). Based on literature of debris removal from roadways, disruptions due to uprooted trees can also be considered in future. Recovery of transportation networks depends on the runoff drainage to nearby lakes and canals.

When a team chooses to restore the next failed component, traffic routes within an area of 14 km radius around the team's current position (e.g., from the failed component just restored) are considered. So, even if a traffic route crosses the boundary of the study area, it was considered in the simulation. A radius of 14 km was chosen because this is the diagonal length of the area containing the transmission lines, towers, and power plants. So, this can be the maximum distance between an origin point and destination point in this area.

Since the restoration team crews can start their work from anywhere like their home or utility office or a temporary hub, the travel time from the initial location of restoration team crews to the first selected component is not considered. In the first time step of the model, restoration teams are readily available at the location of the selected failed component to be repaired. For the next timesteps, the restored components in the previous timestep are the origins for travel time calculation and the components selected to be restored at the current timestep are considered as destinations.

For calculating the crew's travel time, we first added the travel times for each road that makes up the routes between origin-destination (O-D) pairs. The travel time of each road was determined by dividing the road's length by its posted speed limit. Then the shortest path between a given origin-destination pair was found using Dijkstra's algorithm. Finally, we added the travel time of the shortest path to the restoration time. Here, the origin indicates the position of a failed component just restored by a restoration team and the destination indicates the place where the team selects to restore the next failed component. For fuel transportation time, fuel source was considered as the origin and power plant was considered as the destination point. We also considered that traffic lights do not operate without electricity (**Figure 7**).

Transportation network data was collected from USGS National Transportation Dataset (NTD) (sciencebase.gov/catalog/item/5a86ea14e4b00f54eb3a1b55) and Open Street Map (OSM). Traffic light data was collected from Florida Department of Transportation (FDOT).

**Interdependency between Electric Power and Transportation Networks**

Electric power network agents depend on transportation network agents in the following ways. First, transportation networks are needed to transfer fuel like coal, natural gas, and biomass to power plants. We considered that roadways disrupted by runoff cannot transport fuel (Bigger E. John et al. 2009). Fuel will be delivered to power generators via the second-shortest route if the first one is interrupted. Fuel transportation is entirely disrupted till any route reopens. As stated earlier, we assumed that the power plant gets fuel from the nearest source of urban woods. Transportation of fuel to power plant depends on the type of fuel used by the power plant. For example, natural gas and liquid petroleum are transported to power plants via pipelines ("General Pipeline FAQs" 2018) and majority of the coals are usually transported to consumers via railroads ("Coal transported to the U.S. electric power sector declined by 22% in 2020" 2021). Since woody biomass is typically transported to power plants by trucks (Ashton et al. 2016; *Facts about Wood* 2020), we considered roadway for fuel transportation. Second, power restoration teams cannot use disrupted streets. Even if a component has a higher priority (serving more customers), the restoration team cannot fix the failed power component if the roads required to access it are disrupted by flooding. Restoration teams have to repair the components with less priority. This prevents restoring electricity to the greatest number of households in the least amount of time (**Figure 7**).

To capture the dependency of electric power network on transportation network, we considered that failure in the component of the electric power network leads to the loss of traffic signals. Loss of traffic signals is temporarily resolved by following four-way stop signs (Bigger E. John et al. 2009). Although the loss of traffic signals does not entirely block traffic movement on roadways, these disrupted intersections slow down traffic flows (Bigger E. John et al. 2009) and increase the risk of possible crashes (Milian 2017; “Traffic lights out due to power outage from hurricane Ian results in a failure to yield and crash” 2022; Vazquez 2022). However, these negative effects in traffic flow and safety were not simulated in this study.

**Household Agent**

Household agents are used in this study to estimate the power systems resilience to hurricanes. At each time step of the simulation, households’ access to power service is checked.

**Restoration Process of Infrastructures**

*Electric Power Network Agent*

Electricity companies, emergency management agencies, and repair personnel collaborate to restore electricity after a hurricane. The main goal is to safely and quickly restore electricity to as many impacted customers as possible. However, hurricane severity, damage extent, resource availability, and external variables might hamper restoration efforts. This study considered damage assessment, roadway reliance, resource availability, and different restoration strategies.

When the failed components of the electric power network are identified, restoration teams from power companies repair those failed components. Restoration time and resilience depends on restoration strategies (Duffey 2019; Jamal and Hasan 2023b). Thus, restoration priorities affect outage duration. For example, restoration in densely populated regions may be prioritized to serve more households (Esmalian et al. 2022; Liu et al. 2007). Studies found that the duration of power outages is longer in areas with high outage density (high percentage of households without power) (Mitsova et al. 2018). If the hurricane wind speed is low and critical components (substations and transmission towers) do not fail, large populations may not always result in higher outage rates. In such cases, outage density/rate-based priority can improve restoration instead of component based and population based restoration. There is no conventional method for restoring electricity when a severe weather event affects a power network (Esmalian et al. 2022). In this study, we simulated the effect of three restoration strategies:

*(a) Component based restoration*: The ABM first prioritizes repairing of critical components like substation and transmission. Critical components are resource-intensive and serve many consumers. The model then randomly repairs the distribution network including conductors and poles (Esmalian et al. 2022). (see algorithm 1 of SUPPLEMENTAL MATERIALS)

*(b) Distance based restoration*: The ABM first prioritizes repairing of critical components like substation and transmission. Among the transmission components, the ones which are closer to a power plant are prioritized. After the transmission lines are restored, among the distribution lines, the ones which are closer to a substation are prioritized (see algorithm 2 of SUPPLEMENTAL MATERIALS).

*(c) Traffic-light based restoration*: In this strategy, the ABM first focuses on restoring power of traffic lights. Thus, the components supplying power to traffic lights are restored first. The components supplying power to the rest of the households (that did not get back power while restoring the traffic lights) are focused later. In both cases, a specific component is prioritized based on the distance to a power plant (for transmission components) and the distance to a substation (for distribution components) (see algorithm 3 of SUPPLEMENTAL MATERIALS).

*Restoration Scheduling*

**Figure 8** shows the model's dynamic repairing process for power network agents. The process includes several steps: First, based on adopted restoration strategy, priorities are given to power system components. Then for each damaged element, the availability of required resources (i.e. the number of restoration teams) is checked. After that, accessibility to the failed component through roadways is checked. If resources are available and the damaged component is accessible by a repairing team, the failed component will be repaired within the required time as described in agent's section above. If a failed component is not accessible, restoration teams will restore the next prioritized component which is accessible. If no transmission component is accessible but distribution components are accessible, distribution components will be restored before restoring transmission lines. The simulation inputs the number of power restoration teams. Previous studies (Esmalian et al. 2022; Ouyang and Dueñas-Osorio 2014) reported that the number of power restoration teams varies from 800 to 2000 within a week for a county affected by hurricanes (Ouyang and Dueñas-Osorio 2014). Since this study focuses on one ZIP Code and Miami-Dade County consists of 222 ZIP Codes, following the existing literature, this model considers 10 to 12 and 30 to 40 restoration teams for moderate and high wind speed, respectively. If a substation requires more than the mentioned number of teams based on damage status, the model starts with additional number of teams required for the substation. As soon as the substations were restored, additional teams are removed. This can be updated in the simulation by adding the actual number of restoration teams.

*Transportation Network Agent*

Restoration of the components of transportation network depends on the restoration of power network agents (for traffic lights) and runoff drainage rate.

**Quantifying Infrastructure Resilience**

To measure the impact of a disruption on a system, a comprehensive metric is needed. Previous studies adopted different metrics such as the maximum number of outages (Liu et al. 2005), hardship (Esmalian et al. 2022), restoration time, and resilience (Aghababaei and Koliou 2022a; b). Resilience indicates the ability of a system to return to its normal state or situation after a disruption due to an extreme event (Hosseini et al. 2016; Roy et al. 2019). At the onset of a hurricane, system performance starts to decrease in the affected area, reaches maximum drop after a certain time, and then starts to recover. Since system performance is not at usual level, the decrease in system performance compared to regular periods can be termed as the loss of resilience in a disaster affected region (Jamal and Hasan 2023a; Roy et al. 2019). Resilience/loss of resilience considers both the outage rate and the restoration time. Since utility companies aim to restore power service of the largest number customers within the shortest possible time, we considered the loss of resilience to measure the household impact of disrupted infrastructure systems. For a particular hurricane intensity, we did not change the amount of restoration resources. So, the improvement in outage restoration time may not significantly differ under various restoration strategies. However, the restoration rate of households over time can significantly vary under different restoration strategies. To capture this aspect, the loss of resilience is calculated from the ABM results in this study.

We calculated the loss of resilience based on the formula proposed by Bruneau et al. (2003) in the context of infrastructure resilience due to seismic effect (**Equation 6**).

$$TRL_i = \int_{t_0}^{t_1} [1 - Q\ (t)]dt \quad (6)$$

here $TRL_i$ refers to transient resilience loss of an infrastructure system, $i = \{power\ network, transportation\ network\}$. The term "transient" indicates that resilience loss is calculated in response to one disaster event only (Jamal and Hasan 2023a; Roy et al. 2019). $Q(t)$ represents the quality function of a system at a given time $t$ and $(t_1 - t_0)$ is the restoration time. **Figure 9** illustrates these concepts. The region between the horizontal dashed line (representing the baseline value) and reduced quality function curve (solid line) from $t_0$ to $t_1$ defines the transient *loss of resilience* of the system. The horizontal dashed line signifies the expected trajectory of the system in the absence of any disruption. The solid curved line from $t_0$ to $t_1$ indicates the performance of the system impacted by disruptions. The *resilience* is quantified as the area beneath the quality function curve from time $t_0$ to $t_1$ (Jamal and Hasan 2023a).

Different studies have defined different quality function $Q(t)$ depending on the infrastructure considered such as buildings (Abdelhady et al. 2020), bridges (Akiyama et al. 2020) and electric power networks (Aghababaei and Koliou 2022a; b); all are different from one another. We adopted $Q(t)$ following the study by Aghababaei and Koliou on electric power network's resilience measure (Aghababaei and Koliou 2022a; b). The quality function $Q(t)$ for power network is measured by the number of households served at a given time, normalized by the total number of customers. The quality function $Q(t)$ for transportation network is measured by the number of traffic lights functioning at a given time, normalized by the total number of traffic lights.

To understand how much resilience is lost to power service in percentage, we also quantified the $\frac{\text{transient loss of resilience } (\text{TRL}_i)}{\text{maximum possible resilience } (\text{MPR}_i)} \times 100\ \%$. Maximum possible resilience is the amount of resilience under no disruption (baseline condition) within the same time interval (Jamal and Hasan 2023a).

## RESULTS

### Verification and Validation

To perform the verification of the ABM, power network agents were exposed to different hazard intensities to ensure the consistency of their behavior against hazard variations. For instance, a higher wind speed would cause more damages to the components of electric power and transportation networks requiring longer restoration times (see the scenario analysis).

Some validation methods were adopted to ensure the credibility of the ABM outputs. For instance, we checked if the ABM methods to represent infrastructure damages (e.g., fragility curves) and restoration models were previously validated. Furthermore, we compared the ABM outputs in response to different wind speeds against actual observations. For this, the maximum percentage of customers without power and the restoration times available from the ABM were compared to the corresponding actual values from the counties having similar wind speeds during Hurricane Irma in Florida. Since hurricane wind speed varies within a county, for comparison, we chose only the counties with mean wind speeds close to the wind speeds adopted in the ABM simulation (**Table 1**). The simulated restoration times from the ABM in **Table 1** vary within a range since three different restoration strategies are adopted in this study.

**Table 2** presents a list of the major random variables considered in the analysis with explanations given below.

*Wind Speed*

Wind speed is a critical factor in hurricane resilience studies because it directly influences the severity of damage to infrastructure systems. Higher wind speeds increase the likelihood of component failures, such as failed power lines and damaged substations, which in turn affect the duration and extent of power outages.

In this analysis, we selected specific values based on real events to examine the system's response under varying conditions. For instance, we used the exact wind speed recorded during Hurricane Irma (65 mph) to represent actual conditions. Additionally, to assess the system's resilience under a more extreme condition, we considered a higher wind speed of 115 mph (1.8 times the actual wind speed for Irma). This approach allowed us to test the sensitivity of the simulation analysis under different wind speed conditions.

The ASCE 7-16 standard suggests that structures in hurricane-prone regions should be designed to withstand wind speeds between 110 and 140 mph (*Minimum design loads and associated criteria for buildings and other structures* 2017), depending on the location and risk category. Thus, the choice of a specific wind speed value of 115 mph may allow us to explore how system performance is impacted by a wind intensity within this range, particularly in terms of the number of failed components and the necessary restoration efforts.

*Flood Depth*

Flood depth is another critical variable because flooding is a common consequence of hurricanes, especially in coastal and low-lying areas. Floodwater can make roads inaccessible, which hampers the mobility of restoration teams and delays restoration activities. Under this variable, we considered situations such as actual flood depth during Hurricane Irma and no flood situation. We did not consider other flood depths because we assumed a constant floodwater runoff rate. This implies that the time required to clear floodwater from roads would increase linearly from zero to the actual flood depth. Given this linear relationship, analyzing the extreme cases of no flooding and the actual flood depth provides sufficient insights on the impact of flooding on road accessibility and restoration efforts.

*Restoration Strategy*

A restoration strategy refers to the approach used to prioritize and carry out restoration activities. Different strategies (e.g., component based, distance based, traffic-light based) can lead to different outcomes in terms of how quickly power is restored. By comparing different strategies, the study can identify the most effective approaches under various conditions. This helps decision-making on how to optimize restoration efforts in future events.

We evaluate the effect on model outputs exerted by individually varying only one of the model inputs across its entire range of plausible values from **Table 2** (Frey and Patil 2002). Since a systematic sensitivity analysis would require us to run a large number of simulation experiments (Hanh 1972) where each simulation run takes about 20 mins for a lower wind speed and 45 mins for a higher wind speed value, we observe the changes in the overall result with respect to the limited changes in the variables. The sensitivity analysis reveals significant differences in power system's resilience for different values of the variables (see **Table 3**)**.** It is evident that wind speed, flood depth, and restoration strategy are critical factors influencing the power system's resilience. Higher wind speeds and the presence of floodwater on roads lead to significantly greater loss of resilience, underscoring the importance of these variables in our model.

**Scenario Analysis**

The developed ABM was applied to evaluate the following aspects of the restoration of power and transportation networks:

(1) the effect of three types of restoration strategies under different wind speeds,

(2) the effect of road accessibility on power restoration,

(3) the effect of power outage on components of transportation network,

**Effects of Restoration Strategies on Households**

**Figure 10** presents the effect of restoration strategies on power service restoration of households under two different wind speed values. It shows that under both low and high speed values, distance based and traffic-light based restoration strategies perform better than a component based restoration strategy. The restoration curves for distance based and traffic-light based restoration shift to the left (for wind speed 65 mph) and up (for wind speed 115 mph) than component based restoration, indicating that power service of the largest number of customers is restored within the shortest possible time by adopting distance and traffic-light based restoration compared to component based restoration strategy. When a component based restoration strategy is implemented, restoration of the distribution components of power network occurs in a random sequence. This prevents to restore power to the largest number of households within the shortest possible time. On the other hand, when distance based or traffic-light based restoration is adopted, an order is maintained along with the hierarchy between transmission lines and distribution lines. Although the traffic-light based restoration strategy performs better than the distance based restoration strategy when the wind speed is low, they perform in almost similar way when the wind speed is higher. In **Figure 10 (a)**, it drops to 0, because of assuming that fuel cannot be transported to the power plant when roads are disrupted. Initially, all the possible paths between the fuel source and the power plant are not accessible because one or more links/intersections of each path are flooded. After about 16 hours, one of the paths between fuel source and power plant starts functioning because of flood water being cleared from all of its associated links. In **Figure 10 (b)**, under a higher wind speed (115 mph),) the percentage of households with power drops and remains at 0 up to 100 hours since several transmission lines failed which resulted in cascading failures due to the loss of connectivity to power generator. From **Figure 10 (b)**, initial restoration is slow because focus is given on transportation components first and restoration time to repair transmission components is quite long (explained above in transmission agent section).

**Table 4** presents the loss of resilience values over different restoration strategies on households under two different wind speed values. All restoration strategies were simulated considering that restoration teams cannot use any flooded roadway when restoring the failed components**.** It is observed from **Table 4** that, under a lower wind speed value (65 mph), a traffic-light based restoration strategy improves resilience by about 22.5%, but the distance based strategy improves by only 8.6% compared the component based strategy. However, under a higher wind speed value (115 mph), with both restoration strategies, the improvements are around 22% compared to a component based strategy. When the wind speed is increased from 65 mph to 115 mph, there is a significant increase in the loss of resilience irrespective of restoration strategies. For instance, about 30.4% of resilience is lost under a wind speed of 65 mph, whereas about 58% of resilience is lost when the wind speed increases to 115 mph, if a distance based restoration strategy is chosen (TRL/MPR **in Table 4**).

**Table 5** presents the power restoration times of households under different strategies. When wind speed is lower, 75% people get power back in 82 hours under distance based restoration and in 57 hours under traffic-light based restoration. The traffic-light based restoration strategy restored power to more households faster. When wind speed is higher, 75% of people get power back in 326 and 330 hours under distance based restoration and traffic-light based restoration, respectively.

The above results suggest that traffic-light based restoration performs better than distance based restoration under a lower wind speed value, but they perform almost same under a higher wind speed value. The reason behind this result is that fewer poles and conductors fail under a low wind speed and the power components supplying electricity to traffic lights are usually located on major roads, connecting a high number of households in the network. So, just restoring the poles and conductors connected to a traffic light may also allow giving power back to a high number of customers. However, a significant number of poles and conductors may fail under a high wind speed. Thus, even after restoring the poles and conductors connected to traffic lights, the number of remaining failed components can still be high, preventing to restore the power service of many households. The remaining failed components are then restored based on distance based restoration. Thus, under a higher wind speed value the traffic-light based restoration strategy performed almost same as distance-based restoration, but it performed better than distance-based restoration under lower wind speed.

### Effect of Road Accessibility on Power Restoration

Accessible roads indicate that no road is flooded, fuel transportation to the power plant is not disrupted due to flooding, and restoration teams can access all failed components. Whereas inaccessible roads indicate that roads are flooded with varying water depth, fuel transportation to power plant is disrupted due to flooding, and restoration teams cannot use the flooded roads. **Figure 11** shows the effect of road accessibility to restore power services of households under two different wind speed values and when a distance based restoration strategy is adopted. Under both low and high wind speed values, the restoration curves shift to the left when roads are accessible. This indicates that restoration of power services will be faster if restoration teams do not need to wait due to inaccessible roads and fuel transportation to power plants is not delayed. In **Figure 11 (a),** the initial ratio of households with power are different under two scenarios depending on road accessibility to power plant. When roads are inaccessible, besides restoration teams' inability to use the flooded roads, fuel transportation to power plant is disrupted due to flooding leading it to be out of fuel and resulting in the ratio of households with power to be 0. When fuel transportation to power plant is disrupted, all the possible paths between the fuel source and the power plant are not accessible because one or more links/intersections of each path are flooded. After about 16 hours, one of the paths between the fuel source and the power plant starts functioning because of flood water being cleared from all of its associated links (**Figure 11 (a)**). In **Figure 11 (b),** the initial ratio of households with power is 0 under both accessible and inaccessible road conditions and it continues be 0 for more than 16 hours. This occurs because of the connectivity loss to the power plant due to a large number of failed power components, particularly the critical components like transmission towers and lines, under higher wind speed and the additional time required to restore those critical components.

**Table 6** presents how road accessibility affects power network restoration under different wind speeds and when a distance based restoration strategy is adopted. Under low wind speed (65 mph), resilience to power service is lost by 34.32 and 53.16 when roadways are accessible and inaccessible, respectively. The values become 246.08 and 264.19 under a higher wind speed value (115 mph). The reduction in the resilience values due to road inaccessibility is similar (around 18). However, road inaccessibility reduces the resilience of households to power supply by about 55% ($\frac{53.16-34.32}{34.32} \times 100\% = \frac{18.84}{34.32} \times 100\%$) under a low wind speed value, and about 7.5% ($\frac{264.19-246.08}{246.08} \times 100\% = \frac{18.11}{246.08} \times 100\%$)

under a high wind speed value. **Table 6** also indicates longer restoration times when roads are inaccessible. When the wind speed is 65 mph, fewer power components fail resulting in smaller loss of resilience and shorter restoration time. On the other hand, when the wind speed is 115 mph, a lot of power components fail causing power outage to a higher percentage of customers resulting in longer restoration time.

**Effect of Power Outage on Restoration of Transportation Network**

**Figure 12** shows the effect of different restoration strategies to restore power service of transportation network under two different wind speed values. All restoration efforts were simulated considering that restoration teams cannot use any flooded roadways to restore the failed components**.** Under both low and high wind speeds, when the traffic-light based restoration strategy is adopted, both resilience and total restoration time for the traffic lights improved significantly as the restoration curve for traffic-light based strategy significantly shifts to the left compared to the restoration curves for other two strategies. When the wind speed is low (65 mph), both distance based, and component based restoration strategies performed in almost similar way (**Figure 12 (a)**). However, under high wind speed values (115 mph), the distance based restoration strategy outperformed the component based restoration strategy (**Figure 12 (b)**).

**Table 7** presents the effect of restoration strategies on restoration of traffic signals under two different wind speed values. All restoration efforts were simulated considering restoration teams cannot use any flooded roadways to restore a component**. Table 7** suggests that, when the wind speed is low, by adopting distance based restoration, the resilience of traffic lights did not improve compared to component based restoration. However, under low wind speed, traffic-light based restoration improved the resilience by 38.9% compared to component based restoration. Under high wind speed values, distance based, and traffic-light based restoration strategies improved the resilience of traffic lights by 15.7% and 50.9%, respectively, compared to component based restoration. From **Table 7**, when wind speed is lower, all traffic lights get power back in 148 hours under distance based restoration and in 80 hours under traffic-light based restoration. When wind speed is higher, all traffic lights get power back in 408 and 200 hours under distance based restoration and traffic-light based restoration, respectively. Therefore, when traffic-light based restoration is adopted, resilience improves and total restoration time for the traffic signals reduces significantly.

## DISCUSSION

This article presents an agent-based model (ABM) offering a comprehensive modeling framework for assessing infrastructure resilience to wind and flood hazards. The model integrates various agents including hazard, electric power network, transportation network, and household agents, to simulate the dynamic interactions between two infrastructure systems during and after a hurricane. This study adopts a computational simulation framework to represent and assess the effect of interdependent relationships between electric power and transportation infrastructure systems. The integrated agent-based model captures relationships among hazard, risk, restoration strategies, and two infrastructure systems.

Quantifying infrastructure resilience is achieved through the introduction of the transient loss of resilience metric, considering both the outage rate and restoration time. The model's verification and validation processes ensure its reliability and accuracy in capturing the dynamic interactions within a community facing a hurricane.

The ABM produced intuitive results and valuable insights. For example, a higher value of imposed wind speed and disrupted roads are likely to significantly reduce resilience of power systems to hurricanes. Moreover, infrastructure interdependencies and restoration strategies can affect resilience improvement efforts. The ABM simulated the effect of wind speed and different strategies such as component based, distance based, and traffic-light based restoration strategies. When exposed to 65 mph of wind speed, compared to a component based restoration strategy, distance based and traffic-light based restoration strategies improve resilience of households to electricity disruption by 8.6% and 22.5%, respectively. On the other hand, when exposed to 115 mph of wind speed, compared to a component based restoration strategy, distance based and traffic-light based restoration strategies improve household resilience to electricity disruption by 21.9% and 22.5%, respectively. Regarding restoring power service of traffic signals, when exposed to 65 mph and 115 mph of wind speed, a traffic-light based restoration strategy improves the resilience of traffic signals to electricity disruption by 38.9% and 50.9%, respectively than component based restoration efforts. Furthermore, compared to other strategies, a traffic-light based restoration strategy does not significantly increase the restoration time of household power services under low and high wind speeds. These results highlight the importance of considering interdependent relationships among infrastructure systems when designing restoration strategies. A restoration strategy considering the restoration of power service to traffic signals efficiently prioritizes transportation network recovery without any loss of household resilience to power service or increase in their restoration time.

This study extends the current literature on utilizing agent-based modeling (ABM) for strengthening infrastructure resilience against power outage. Previous studies used ABM to assess household resilience against power service disruptions (Esmalian et al. 2022) and how resilience to power service decreases because of its dependence on water infrastructures (Aghababaei and Koliou 2022a; b). On the other hand, considering the interdependent relationships between power and transportation infrastructure systems, we propose restoration strategies so that traffic signals receive power without delaying the restoration time to households' power service. The insights gained from this ABM can inform proactive strategies for hurricane preparedness and restoration efforts, ultimately contributing to the development of more resilient and sustainable infrastructure systems. Given the availability of similar data, the model can assist in resilience evaluation in a variety of scenarios.

This study has several limitations which can be addressed in future research. For instance, future research can consider the effect of disruptions due to treefall induced debris on roadways on power service restoration. Increased delays associated with specific traffic conditions such as traffic congestion or crashes at disrupted intersections were not considered in this study. As a result, the estimated travel times could be optimistic, particularly in scenarios where such delays are significant. Future studies may consider using real-world travel times using APIs from services like Google Maps or Waze which aggregate data from a large number of users and provide up-to-date travel times. Thus, it can better account for slow traffic flow and possible crashes to provide more accurate travel time estimates. The generalizability of the ABM results can be assessed by implementing the methodology across the power and transportation infrastructure systems of other hurricane prone areas and/or larger areas (multiple ZIP Codes). The ABM can be extended including other infrastructure systems such as water/wastewater distribution and communication systems.

## CONCLUSIONS

This paper presents an agent-based model (ABM) aimed at modeling infrastructure resilience to hurricanes, focusing on the interdependency between electric power and transportation networks. To represent the dynamic interactions that occur within a community both during and in the aftermath of a hurricane, relationships between two infrastructure systems, hazards, and restoration strategies are incorporated into the integrated ABM. Interdependencies between the two networks are represented in two ways: (i) fuel delivery to power plants and access for restoration teams depend on transportation network, and (ii) through the impact of power outage on transportation network components. The ABM is implemented considering the household location and infrastructures of a community (ZIP Code 33147) in Miami-Dade County, Florida. We simulate three different restoration strategies: component based, distance based, and traffic-light based restoration. Scenario analyses investigate the influence of restoration strategies, road accessibility, and wind speed intensities on power restoration. The results suggest that a traffic-light based restoration strategy efficiently prioritized transportation network recovery while reducing household power restoration time and enhancing resilience. Besides, it was found that restoration of power services will be faster if restoration teams do not need to wait due to inaccessible roads and fuel transportation to power plants is not delayed. As such, interdependent relationships among infrastructure systems can significantly affect the performance of a restoration strategy.

The developed agent-based model can be used as a decision-support tool by policy makers and utility/emergency managers to assess different what-if scenarios for power outage restoration using available resources. By utilizing an integrated agent-based simulation tool, researchers can generate simulated data to develop and test interdisciplinary theories on disaster resilience and investigate complex phenomena that are beyond the scope of observational data (Esmalian et al. 2022; Reilly et al. 2021).

Besides, this tool can be utilized by utility companies and emergency management agencies to assess resilience and restoration time of power and transportation networks, and in decision-making such as which restoration strategy would be beneficial and how many restoration teams should be employed to restore power service to households and traffic intersections. In future, this tool and methodology can be extended by considering other infrastructures (e.g., water supply, wastewater, communication systems) and tested over different study areas. Also, fragility curves for other natural hazards can be included to develop a more generalizable simulation model.

## DATA AVAILABILITY STATEMENT

Some or all data, models, or code that support the findings of this study are available from the corresponding author upon reasonable request.

## ACKNOWLEDGMENTS

The research is funded by the U.S. National Science Foundation through the grants CMMI 1832578, 1832680, and 1832693. The authors are solely responsible for the findings presented here.

## SUPPLEMENTAL MATERIALS

Algorithms 1–3 are available online in the ASCE Library (www.ascelibrary.org).

# Figures

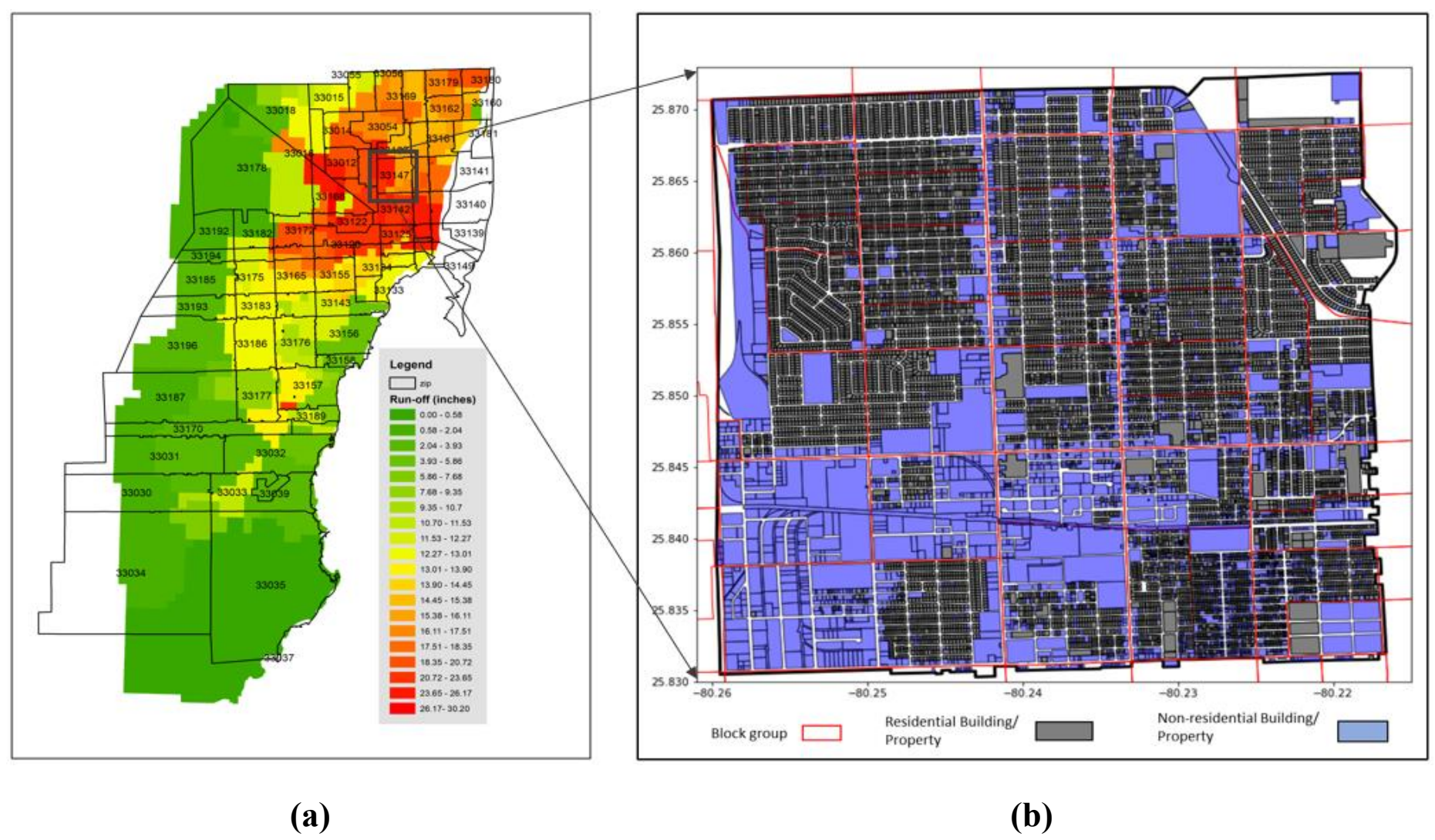

**(a)** **(b)**

**Figure 1 Study Area. (a) Run-off depth distribution in Miami-Dade County (b) Buildings over the block-group of ZIP Code 33147**

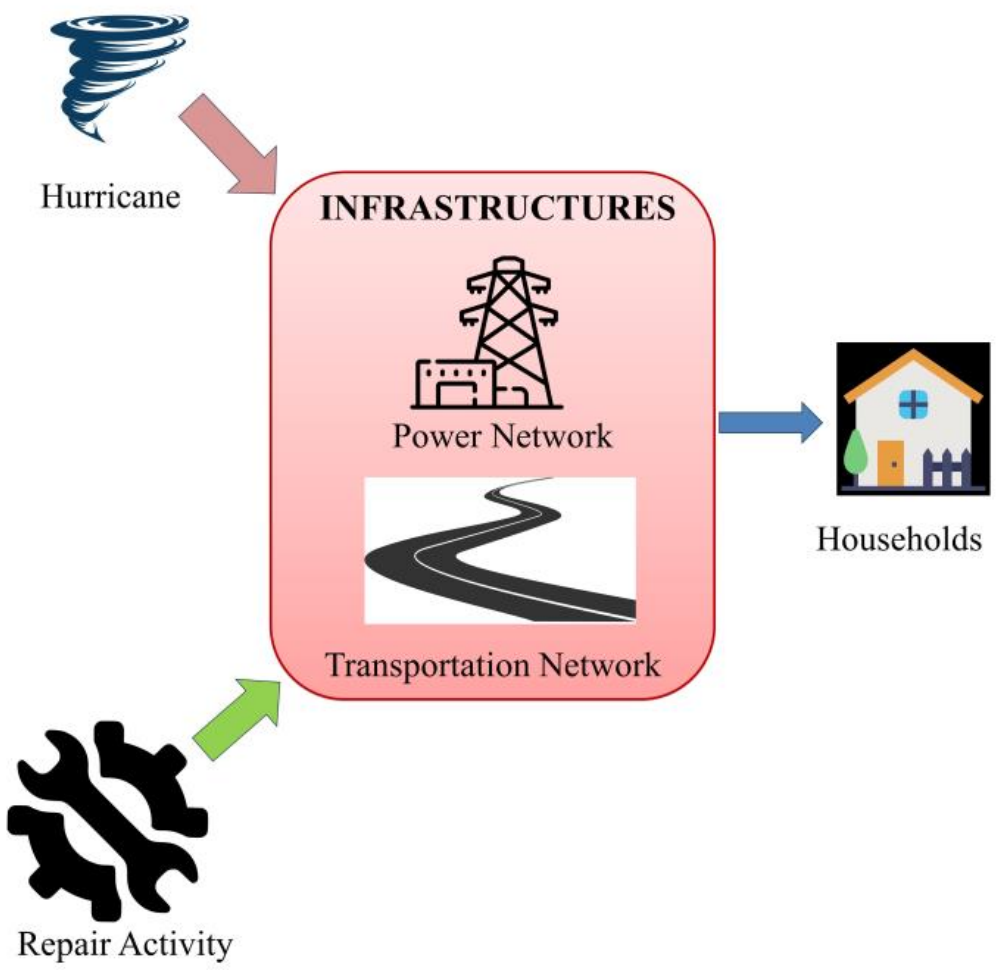

**Figure 2 Overall framework of the dynamic processes in ABM**

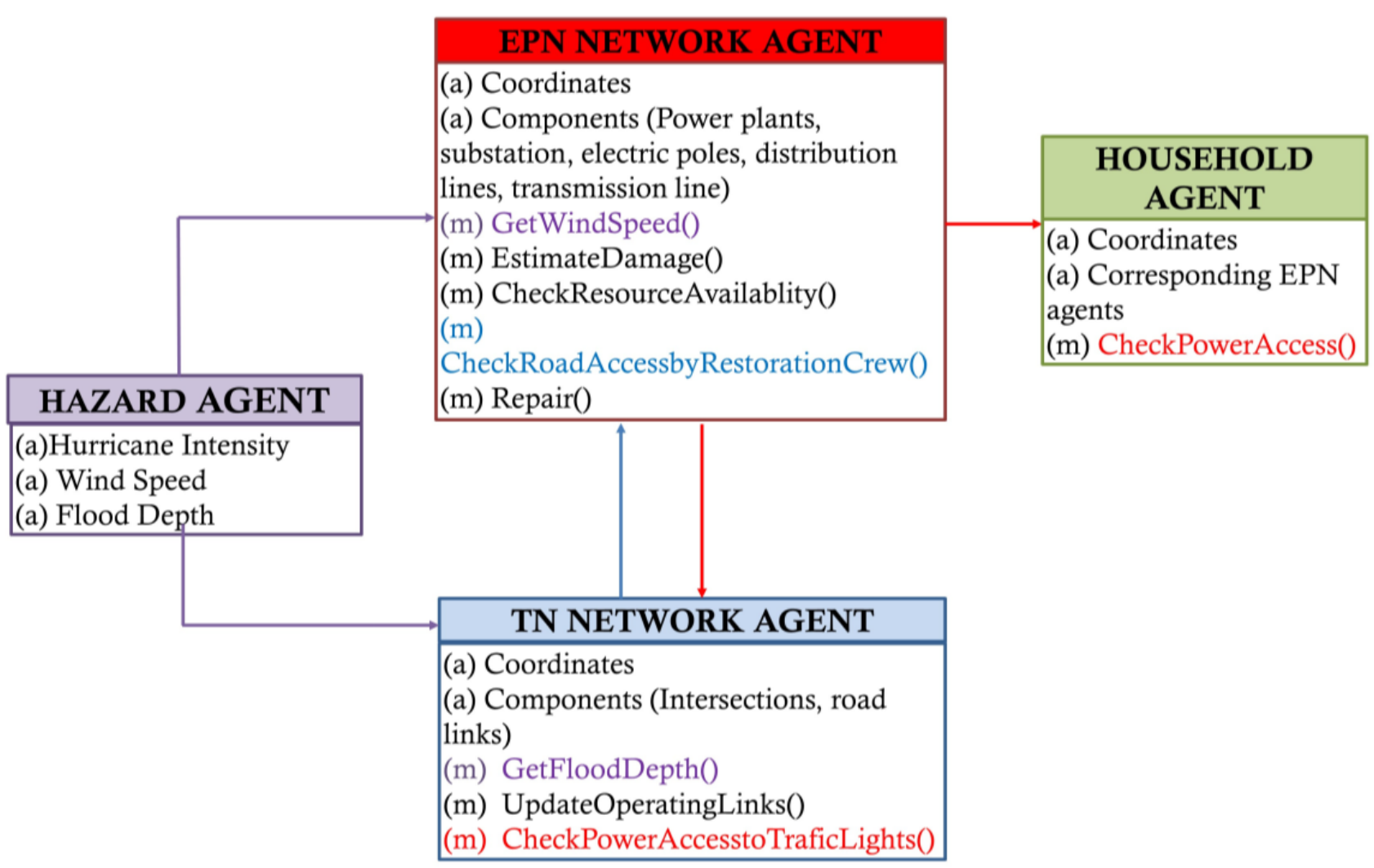

**Figure 3 Unified modeling language class diagram of the agent-based model** [here (a) stands for attributes and (m) stands for methods]

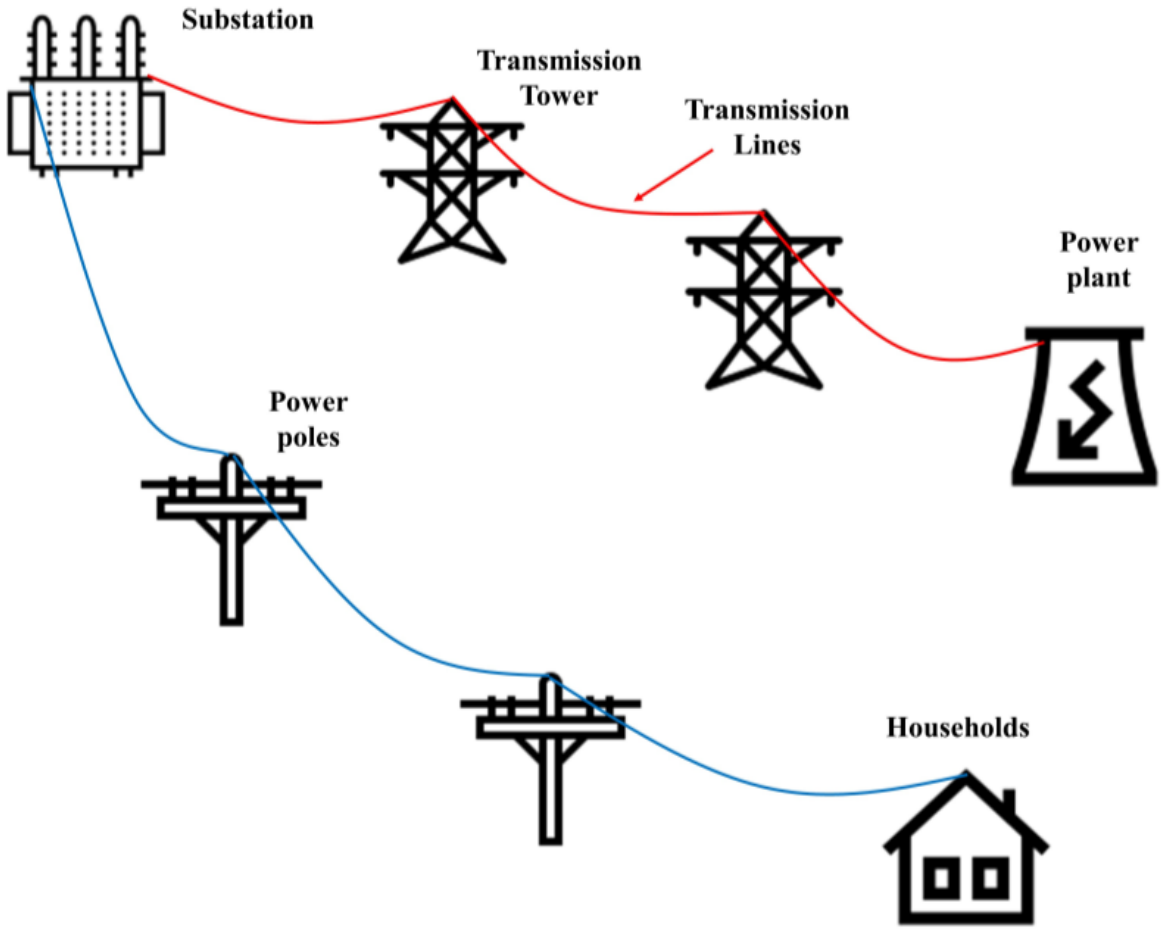

**Figure 4 Electricity flow from a power plant to households**

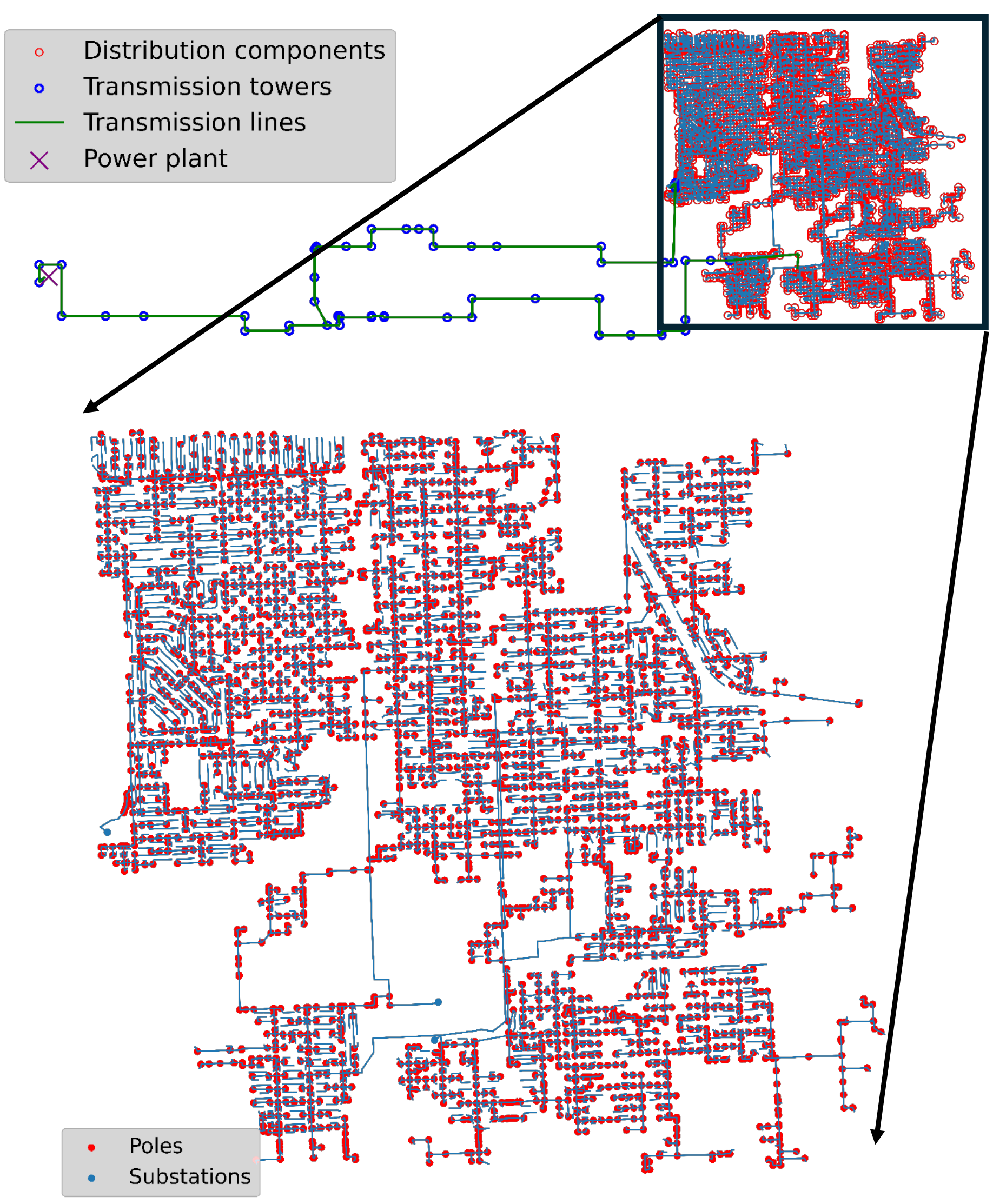

**Figure 5 Power network agents for the study area**

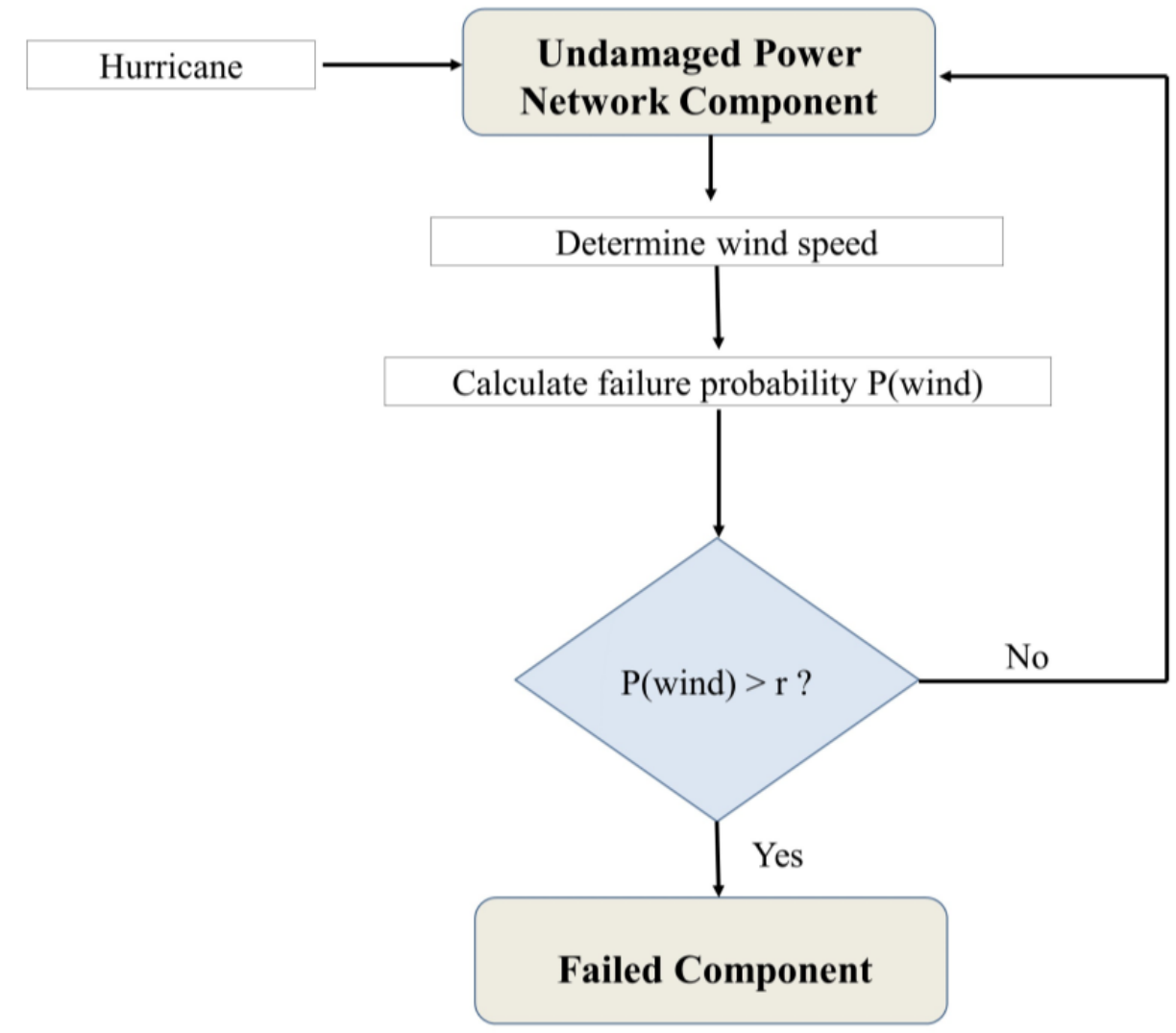

**Figure 6 Schematic overview for power system failure**

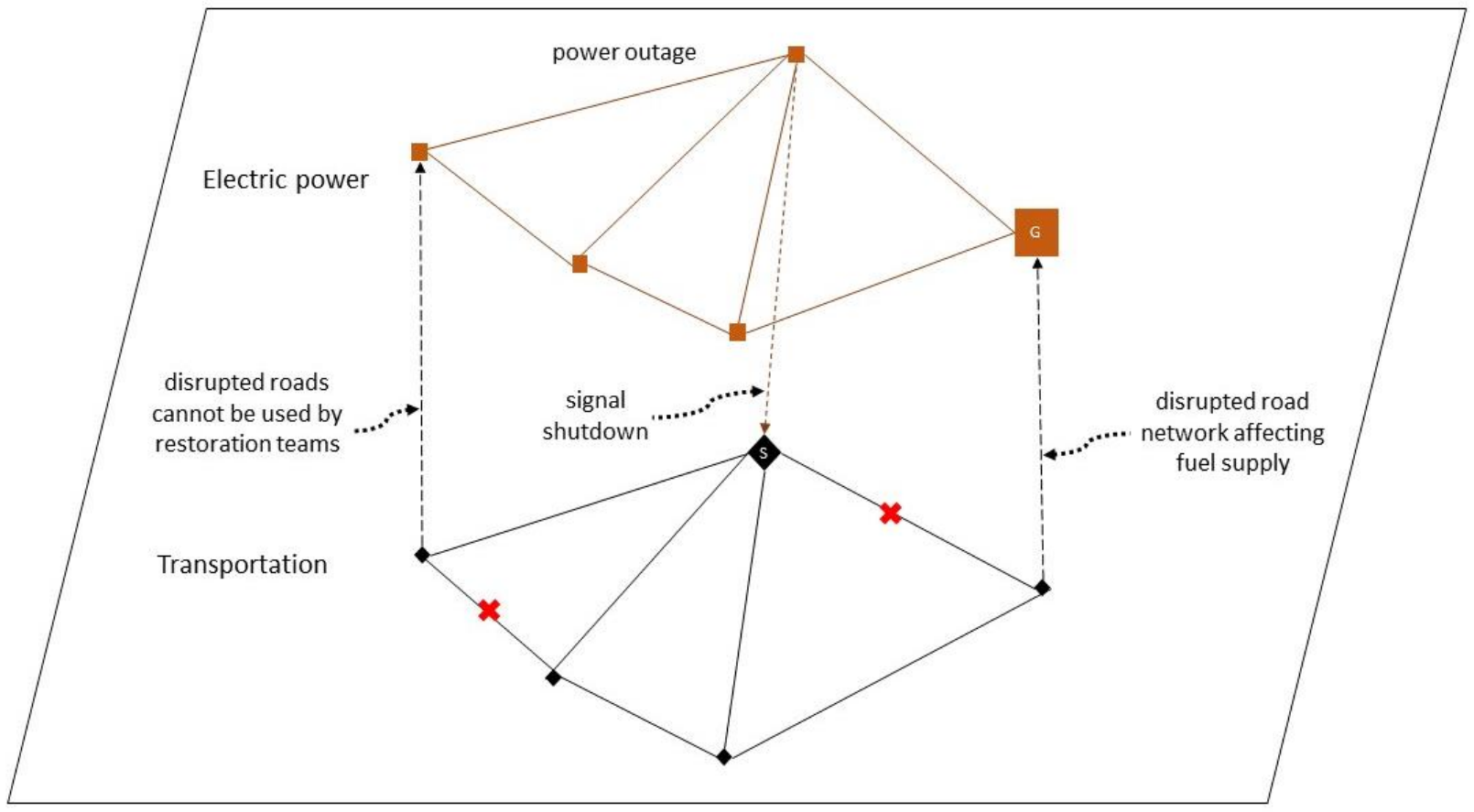

**Figure 7 Interdependent electric power and transportation network systems**

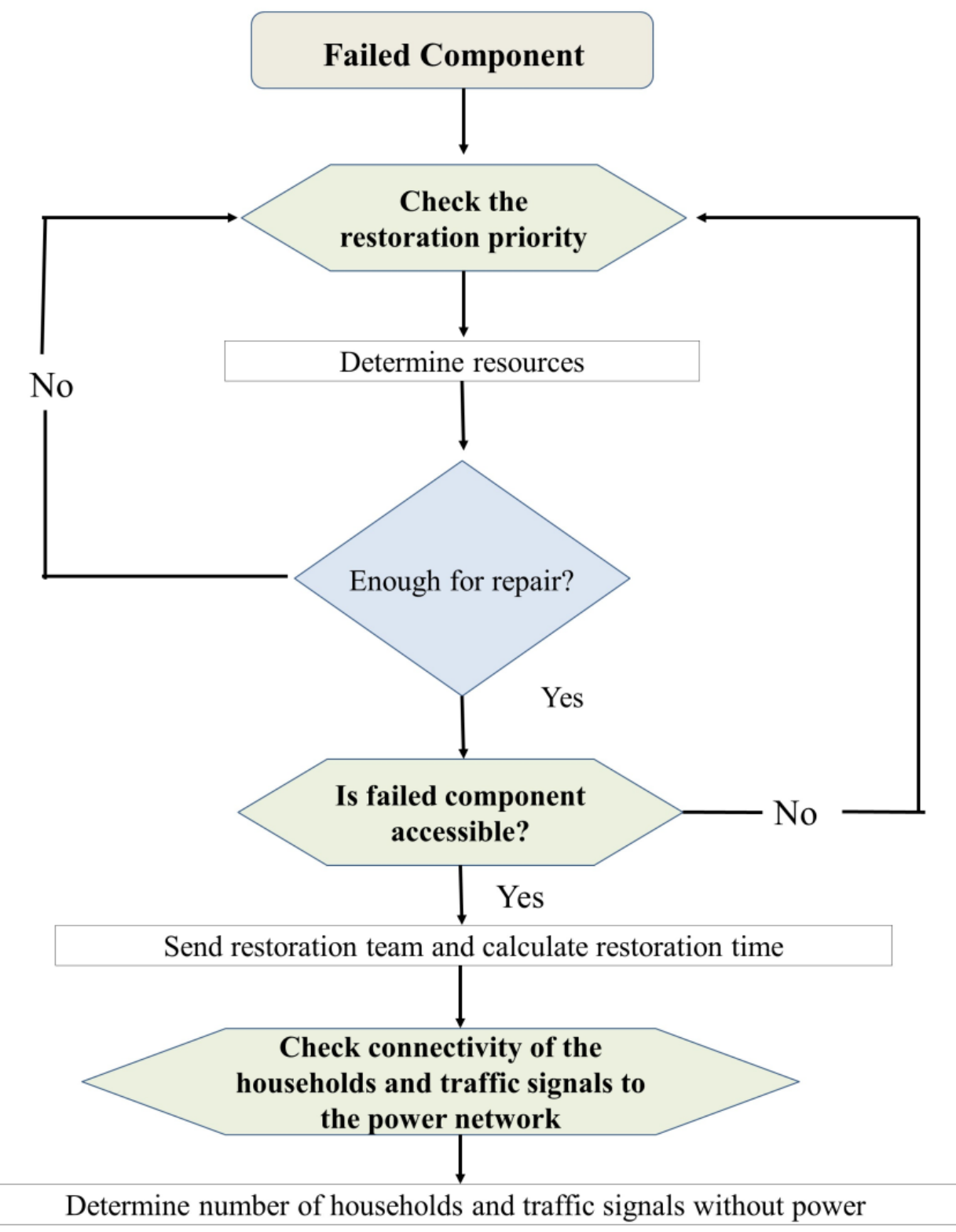

**Figure 8 Flowchart for restoration**

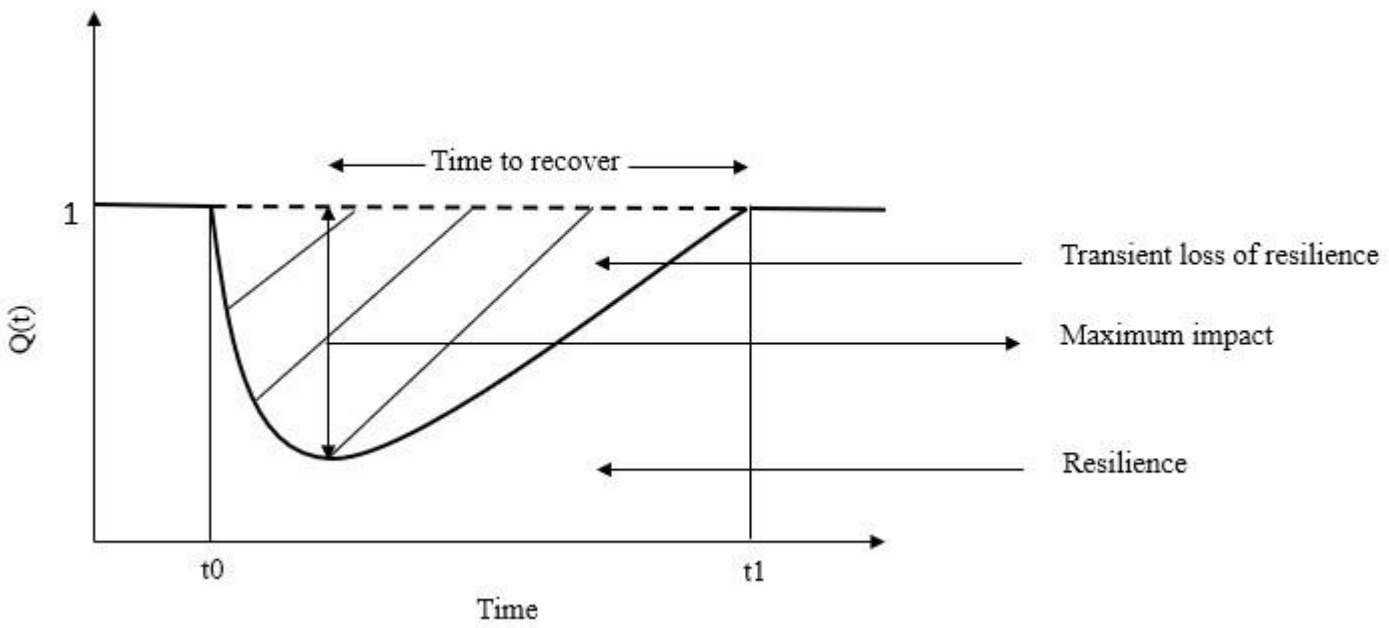

**Figure 9 Resilience and transient loss of resilience**

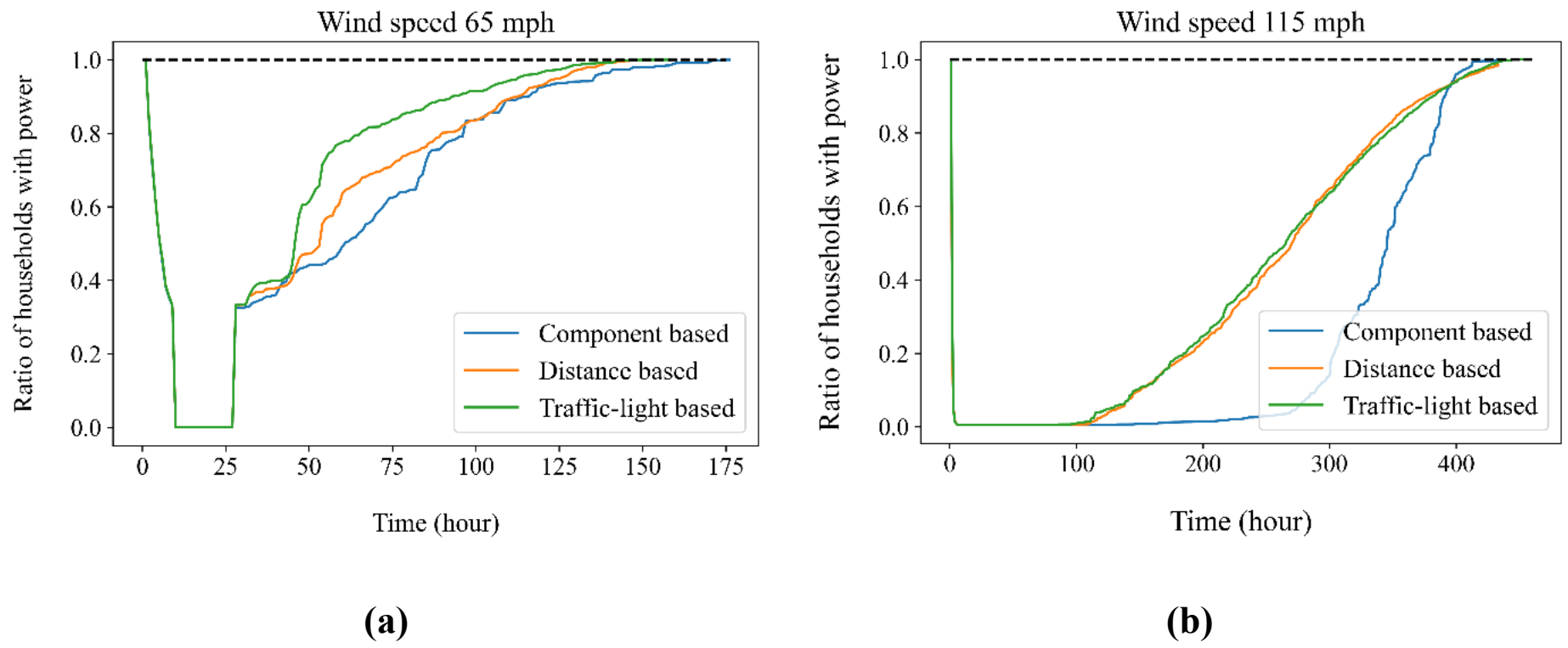

**Figure 10 Effect of different restoration strategies over time**

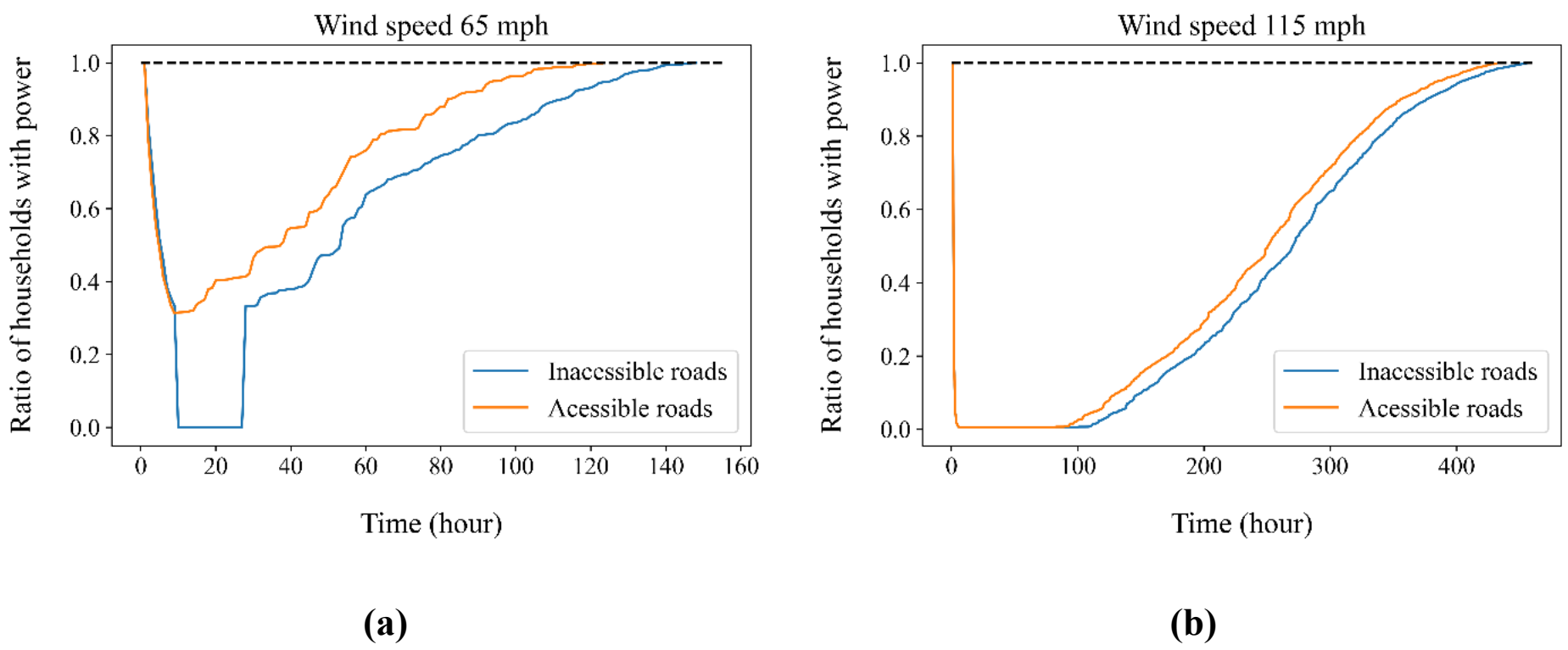

**Figure 11 Effect of road accessibility on power restoration**

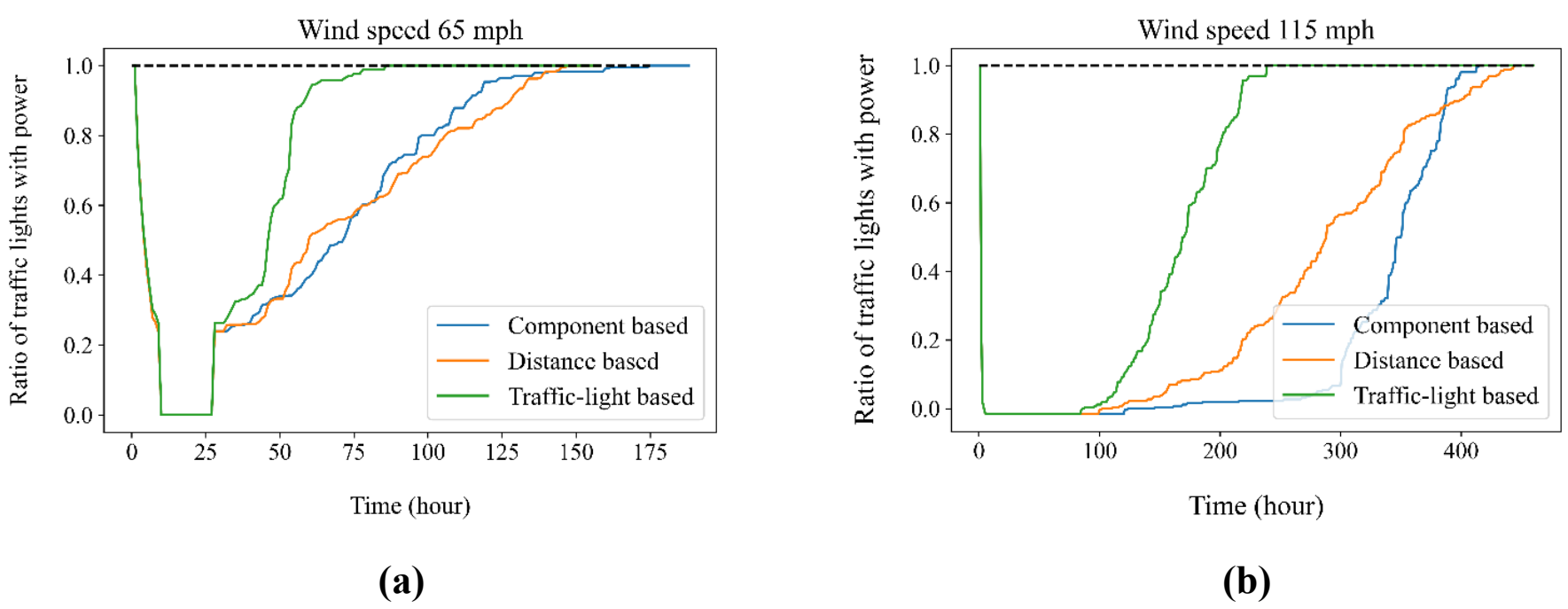

**Figure 12 Effect of power outage on components of transportation network**

## Tables

**Table 1 Comparison between outputs from the ABM and actual values**

| Wind Speed (mph) | *% of customers without power* (ABM) | % of customers without power (Actual) | Restoration time in days (ABM) | Restoration time in days (Actual) |
|---|---|---|---|---|
| 65 | 66 | Miami-Dade =72, Hernando = 61, Flagler = 70 | 6 - 7 | Miami-Dade =6, Hernando = 5, Flagler = 7 |
| 115 | 98 | Hardee = 97, Collier = 86, Hendry = 100, Highlands = 99 | 17.5 -19 | Hardee = 10, Collier = 15, Hendry = 17, Highlands = 12 |
| *This value is obtained without considering the power plant's out of fuel situation. | | | | |

**Table 2 List of variables**

| **Variables** | **Values considered** |
|---|---|
| Wind speed (mph) | 65, 115 |
| Flood depth (inch)* | 0, 1.26 – 30.20 |
| Restoration strategy | Component based, Distance based, Traffic-light based |
| *When flooded, flood depths were different across the study area ranging from 1.26 inches to 30.20 inches with an average depth of 15.72 inches. | |

**Table 3 Sensitivity analysis results ($TRL_{power}$ values with standard deviation in parentheses are reported)**

| | | Component based restoration strategy | | Distance based restoration strategy | | Traffic-light based restoration strategy | |
|---|---|---|---|---|---|---|---|
| | | **Wind speed (mph)** | | | | | |
| | | **65** | **115** | **65** | **115** | **65** | **115** |
| **Average flood depth (inch)** | **0** | 39.52 (4.7) | 320.04 (24.8) | 34.32 (4.2) | 246.08 (22.7) | 25.97 (4.1) | 243.68 (22.5) |
| | **15.72** | 58.18 (7.3) | 338.23 (25.1) | 53.16 (6.1) | 264.19 (23.9) | 45.07 (5.3) | 262.10 (25.6) |

**Table 4 Loss of resilience for households towards power service**

| Restoration strategy | Wind speed (mph) | Transient Resilience Loss ($TRL_{power}$) (ratio * hour) [Standard Deviation] | ($TRL_{power}/MPR_{power}$) *100%[1] | Improvement in resilience (%)[2] |
|---|---|---|---|---|
| Component based | 65 | 58.18 (7.3) | 33.2 | - |
| Distance based | 65 | 53.16 (6.1) | 30.4 | 8.6 |
| Traffic-light based | 65 | 45.07 (5.3) | 25.8 | 22.5 |
| Component based | 115 | 338.23 (25.1) | 73.5 | - |
| Distance based | 115 | 264.19 (23.9) | 57.4 | 21.9 |
| Traffic-light based | 115 | 262.10 (25.6) | 57.0 | 22.5 |

[1] Maximum Possible Resilience (MPR) = 174 (for wind speed of 65 mph) or 456 (for wind speed of 115 mph). This is determined based on a previous study (Jamal and Hasan 2023a).

[2] Improvement in resilience is compared to component based restoration under same scenario.

**Table 5 Restoration time of households (HHs) to power service**

| Restoration strategy | Wind speed (mph) | Restoration time for 75% HHs (hour) [Standard Deviation] | Restoration time for 90% HHs (hour) [Standard Deviation] | Restoration time for 100% HHs (hour) [Standard Deviation] |
|---|---|---|---|---|
| Component based | 65 | 87 (15.5) | 115 (18.6) | 175 (10.6) |
| Distance based | 65 | 82 (9.6) | 113 (11.5) | 148 (10.2) |
| Traffic-light based | 65 | 57 (8.0) | 93 (12.0) | 151 (9.3) |
| Component based | 115 | 380 (21.1) | 392 (19.1) | 450 (20.9) |
| Distance based | 115 | 326 (32.1) | 377 (37.2) | 456 (39.2) |
| Traffic-light based | 115 | 330 (28.4) | 382 (29.0) | 445 (30.2) |

**Table 6 Loss of resilience and restoration time of households (HHs) considering road accessibility**

| Road accessibility | Wind speed (mph) | Transient Resilience Loss ($TRL_{power}$) (ratio * hour) [Standard Deviation] | Reduction in resilience (%) | ($TRL_{power}$/ $MPR_{power}$) *100% | Restoration time for 75% HHs (hour) | Restoration time for 90% HHs (hour) | Restoration time for 100% HHs (hour) |
|---|---|---|---|---|---|---|---|
| Yes | 65 | 34.32 (4.2) | - | 19.6 | 60 | 98 | 122 |
| No | 65 | 53.16 (6.1) | 55 | 30.4 | 90 | 126 | 148 |

| Yes | 115 | 246.08 (22.7) | - | 53.5 | 285 | 360 | 433 |
|---|---|---|---|---|---|---|---|
| No | 115 | 264.19 (23.9) | 7.5 | 57.4 | 305 | 380 | 456 |
| MPR= Maximum Possible Resilience | | | | | | | |

**Table 7 Loss of resilience and restoration time of traffic signals towards power service**

| Restoration strategy | Wind speed (mph) | Transient Resilience Loss ($TRL_{trans}$) (ratio * hour) [Standard Deviation] | ($TRL_{trans}/MPR_{trans}$)[1] *100% | Improvement in resilience (%)[2] | Restoration time for 100% traffic signals (hour) |
|---|---|---|---|---|---|
| Component based | 65 | 64.35 (13) | 36.7 | - | 165 |
| Distance based | 65 | 65.60 (11) | 37.5 | -1.9 | 148 |
| Traffic-light based | 65 | 39.31 (5.6) | 22.5 | 38.9 | 80 |
| Component based | 115 | 344.02 (20.84) | 74.8 | - | 395 |
| Distance based | 115 | 290.11 (23.5) | 63.1 | 15.7 | 408 |
| Traffic-light based | 115 | 169.06 (22.3) | 36.8 | 50.9 | 200 |
| [1]Maximum Possible Resilience (MPR) = 174 (for wind speed of 65 mph) or 456 (for wind speed of 115 mph) [2]Improvement in resilience is compared to component based restoration under the same scenario. | | | | | |